\documentclass[%
 reprint,
 amsmath,amssymb,
 aps,
]{revtex4-2}

\usepackage{graphicx}% Include figure files
\usepackage{dcolumn}% Align table columns on decimal point
\usepackage{bm}% bold math
\begin{document}

\preprint{APS/123-QED}

\title{Permutation entropy of indexed ensembles: Quantifying thermalization dynamics}

\author{Andr\'es Aragoneses}
\affiliation{%
Department of Physics, Eastern Washington University, Cheney 99004, WA, USA
}
\author{Arie Kapulkin}
\affiliation{%
Department of Physics and Astronomy, Carleton College, Northfield 55057, MN, USA
}
\author{Arjendu K. Pattanayak}
\affiliation{%
Department of Physics and Astronomy, Carleton College, Northfield 55057, MN, USA
}

\date{\today}% It is always \today, today,

\begin{abstract}
We introduce `PI-Entropy' $\Pi(\tilde{\rho})$ (the Permutation entropy of an Indexed ensemble) to quantify mixing due to complex dynamics for an ensemble $\rho$ of different initial states evolving under identical dynamics. We find that $\Pi(\tilde{\rho})$ acts as an excellent proxy for the thermodynamic entropy $S(\rho)$ but is much more computationally efficient. We study 1-D and 2-D iterative maps and find that $\Pi(\tilde{\rho})$ dynamics distinguish a variety of system time scales and track global loss of information as the ensemble relaxes to equilibrium. There is a universal S-shaped relaxation to equilibrium for generally chaotic systems, and this relaxation is characterized by a \emph{shuffling} timescale that correlates with the system's Lyapunov exponent. For the Chirikov Standard Map, a system with a mixed phase space where the chaos grows with nonlinear kick strength $K$, we find that for high $K$, $\Pi(\tilde{\rho})$ behaves like the uniformly hyperbolic 2-D Cat Map. For low $K$ we see periodic behavior with a relaxation envelope resembling those of the chaotic regime, but with frequencies that depend on the size and location of the initial ensemble in the mixed phase space as well as $K$.  We discuss how $\Pi(\tilde{\rho})$ adapts to experimental work and its general utility in quantifying how complex systems change from a low entropy to a high entropy state.
\end{abstract}

\maketitle

%\tableofcontents

\begin{figure*}[ht]
\centerline{
\includegraphics[width=14.5cm]{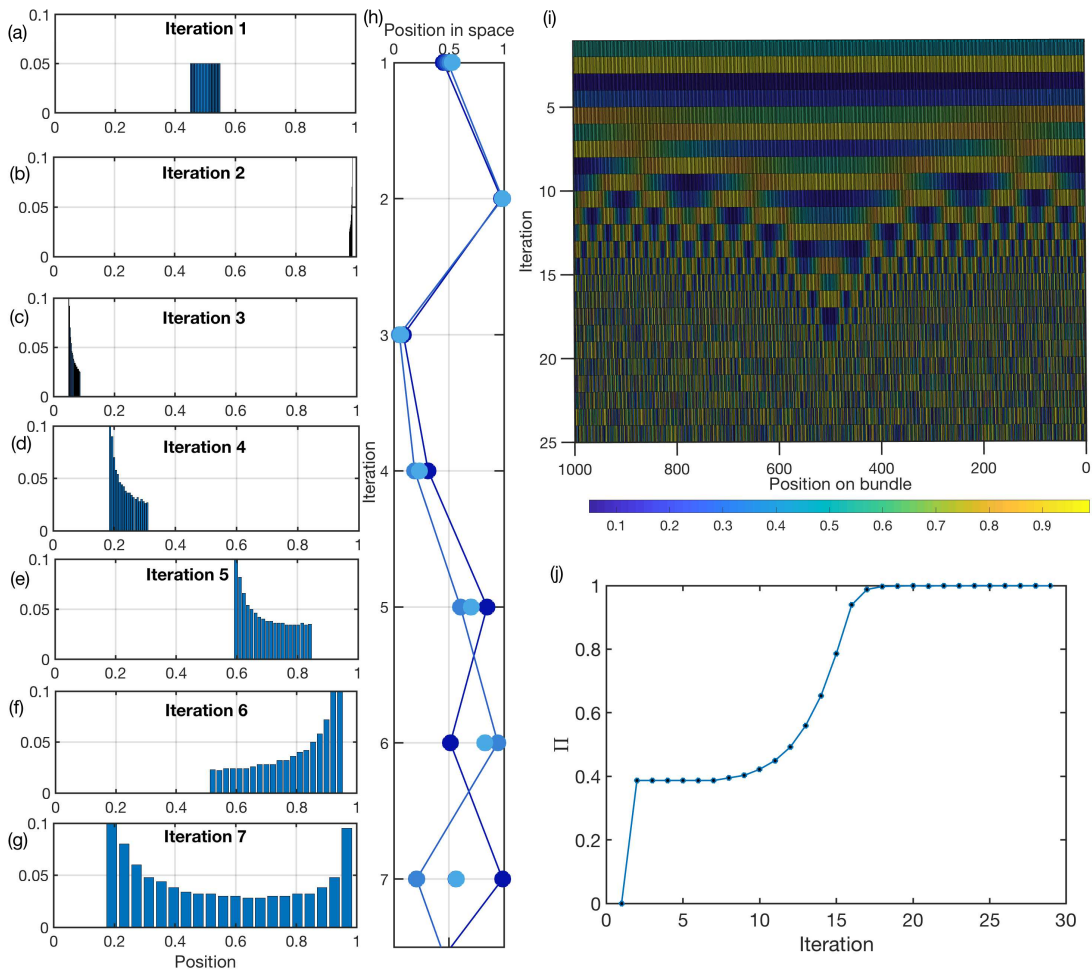}
}
\caption{Evolution of an ensemble for the logistic map with $r=3.95$ given 1000 members equally spaced across position space ($\delta x_1=x_{1,{n+1}}-x_{1,n}=10^{-4},~\forall n$) and initially localized to $0.45\leq x_1^n\leq 0.55$. (a)-(g) Position space probability distribution function for the ensemble. The ensemble starts in a narrow region and spreads out with time. (h) The first few iterations for a chosen set of ensemble members using dark blue for elements initially closer to $x=0.45$ ranging to light blue for those initially closer to $x=0.55$. Lines indicate elements $x_1^{1}=0.45$ and $x_1^{400}=0.49$. The initial ensemble expands and the trajectories of the elements entangle. (i) Landscape of the temporal evolution of the ensemble. Color code indicates the position of each element in space (blue for $x_{i,n}=0$ to yellow for $x_{i,n}=1$). Elements of the ensemble are distributed in the horizontal axis, and time in the vertical axis, from top to bottom. Regular structure is manifest for initial iterations that are washed out as the system evolves. (j) $\Pi$-Entropy, $\Pi$, of the ensemble as a function of time. PI-entropy captures the internal structure among the elements of the ensemble. The ensemble starts ordered, $\Pi=0$, but as the system evolves the elements intertwine and entropy increases. Words of $D=3$ are used to compute $\Pi$.}
\label{ensemble}
\end{figure*}

Far-from-equilibrium ensembles or probability densities $\rho(t)$ describe a variety of phenomena involving matter, energy, or information transport in fundamental physical, chemical and biophysical systems. The entropy dynamics of $\rho(t)$ fully characterizes the approach to equilibrium, the type of the equilibrium (meta-stable, unstable, steady-state), and the variety of solutions possible in any given system, but are intractably difficult to analyze far from equilibrium. The alternate program to characterize entropy dynamics via formal $\rho$ dynamics\cite{1994_Zurek_PRL,1997_PRL_Pattanayak,1999_PRL_Pattanayak} arising from interest in the quantum limit, or via an ensemble $\rho$ constructed from individual trajectories\cite{1999_PRL_Latora, 2002_PLA_Latora, Falcioni_05, 2004_PRL_Tsallis,2017_JSMTE_Tsallis} established a connection between thermodynamic entropy growth and the dynamical loss of information about trajectories. Recent discussions use ensemble dynamics to help understand systems with parameter drift as well as to as snapshot techniques to capture the shape of invariant distributions\cite{2021_NLD_Janosi, 2022_PRE_Janosi}. However, progress is hampered since calculating $\rho$ dynamics, either using many individual trajectories OR by propagating partial differential equations, prove computationally challenging. Ironically, accurate $\rho$ dynamics require very fine grained calculations but for Hamiltonian evolution coarse-graining (smoothing over fine scales) is necessary for a time-dependent entropy. 

Inspired by Permutation Entropy~\cite{2002_PRL_BP, 2022_EPL_Zanin} (PE) used for time-series, we propose `PI-Entropy' (the Permutation entropy of an Indexed ensemble) $\Pi(\tilde{\rho})$ which quantifies the shuffling of neighboring ensemble elements as a measure that connects thermodynamic entropy with the mixing and folding due to complex trajectories for ensemble members. The use of indexed ensembles $\tilde{\rho}$ and the focus on `digitised' shuffling proves to be extremely computationally efficient relative to calculating $\rho$ itself. We are able to use minimal computational effort to explore the approach to equilibrium for the Logistic and other 1-D maps, as well as the 2-D Cat and Chirikov Standard Maps, in the latter case accessing previously unexplored mixed phase space regimes. In particular, just as the PE is strongly correlated with the Lyapunov exponent $\lambda$~\cite{1978_BBMS_Ruelle,1977_RMS_Pesin,2014_JMP_Ohya,2020_Entropy_Keller, 2018_Entropy_Trostel}, $\Pi$-Entropy acts as an excellent proxy for the \emph{change} in the coarse-grained thermodynamic entropy. We find that $\Pi$ dynamics are intuitive and reproduce previous results, as well as provide new insights. 

Specifically: (1) For `elementary' chaotic systems in 1-D or 2-D, which have little initial condition dependence, $\Pi$ relaxes to equilibrium with a smooth and universal S-shape which allows us to define a time scale 1/$\alpha$ for relaxation to equilibrium. We find that $\alpha$ varies monotonically with ensemble-averaged versions of $\lambda$ and PE. (2) The mixed (and hence highly initial condition dependent) phase space of the Standard Map yields more complex dynamics: For sufficiently large nonlinearity $K$ where the phase space is almost entirely chaotic, the $\Pi$-Entropy evolves as for the Cat Map. For lower $K$ we have mixed phase spaces and $\Pi$ oscillates with an overall envelope that resembles chaotic systems. The oscillation frequencies depend on $K$ as well as on the details of the initial support of the ensemble, and we identify different internal  time scales for the mixed-phase space and integrable regime. Given this ease of use including in experiments, $\Pi(\tilde{\rho})$ is a promising approach to quantify complex non-equilibrium ensemble dynamics. 

In the following, we introduce $\Pi$ in the context of the Logistic Map, before moving to other 1-D maps. We demonstrate universality in the relaxation to equilibrium for these dynamics via a $\Pi, \dot{\Pi}$ Entropy Phase Space (EPS), before moving to the Hamiltonian 2-D uniformly hyperbolic Cat Map and finally the Standard Map. We conclude with a short discussion, including the prospects for using this experimentally viable technique elsewhere. Consider an ensemble of $N$ trajectories each
%consisting of N elements (each representing an individual complex dynamical system) 
evolving according to the logistic map 
\begin{equation}
    x_{i+1}^n = rx_i^n(1-x_i^n) ~,
\end{equation}
where $r$ is a parameter controlling the system dynamics, $n=1, 2, 3, ..., N$ labels ensemble elements, and $i=1, 2, 3, ...$ denotes discretized time. This ensemble $\rho$ occupies a phase space neighborhood, and each $x_1^n$ is understood to be sampled from a $\rho(x)$ itself governed by the corresponding Frobenious-Perron operator.
We aim to quantify the stretching and folding that $\rho$ undergoes during complex dynamics, leading to the loss of correlation between trajectories of initially close ensemble members. This loss of information is quantified at the trajectory level by the Lyapunov exponents or the Kolmogorov-Sinai entropy for chaotic systems.

Figure~\ref{ensemble} shows such evolution for an ensemble with details as given in the caption. The evolution of $\rho$ over the first seven iterations are seen in Figs.~\ref{ensemble}(a-g). The ensemble initially remains compact while moving through phase space. Then follows a stage where $\rho$ spreads out and relaxes to an invariant distribution covering the entire phase space (while individual elements continue to evolve). This intuitive visualization of $\rho$ does not show how the elements move \emph{relative} to each other, i.e. how correlations evolve. In Figure~\ref{ensemble}(h) we shift attention to the indexed ensemble $\tilde{\rho}$. Here we see (from top to bottom) a few of the tracked indexed trajectories, demonstrating how they braid across each other in position space. This loss of dynamical correlation between initially neighboring points is reflected in the growth of fine-grained phase space structure at a rate given by generalized Lyapunov exponents\cite{1997_PRE_Pattanayak}.
%In this figure we see that not only do elements of the ensemble spread out and otherwise change the distribution until -- even though the trajectories are evolving -- there is a final invariant distribution, but also that these trajectories braid across each other in position space, as expected from the complexity of the dynamics.
%We can look at this same evolution from a different perspective as we show in
Figure~\ref{ensemble}(i) presents the evolution of the corresponding indexed ensemble $\tilde{\rho}$ where time $i=0-25$ is along the vertical axis, the ensemble indices $n=1-1000$ are on the horizontal axis, and the position $x$ is shown as a color ($blue=0 < x_{i}^n < 1=yellow$). Reading down vertically, the initial ($i=1$) narrowly localized ensemble is all green. As shown for $\rho$ in Figs.~\ref{ensemble}(a-g), $\tilde{\rho}$ moves to the right (yellow), and then to the left (blue) while remaining localized. $\tilde{\rho}$ then spreads and, as it spans the dynamical inflection point at $x=0.5$, the trajectory histories start folding over and braiding together, as is apparent in the growing range of colors at each iteration. The initial uniform $\tilde{\rho}$ evolves to transient intermediate states with structure at increasingly finer scales structures until it reaches a different near-uniformity, of being too fine grained to be discernible. As expected for chaos, the initially ordered $\tilde{\rho}$ has become featureless, and correlations with neighbors have disappeared, taking the system from an ordered state to a highly disordered state. Notably, these dynamics for $\tilde{\rho}$ are visible to finer length scales and hence on a far longer time scale than are visible in Figs.~\ref{ensemble}(a-g) for $\rho$ itself. \\
%Note: We should probably calculate the von-Neuman entropy of the distribution as well.

We quantify the loss of correlations using techniques inspired by Permutation Entropy (PE)~\cite{2002_PRL_BP}. The PE technique discretely samples a dynamical time series and uses the relative populations of ordinal patterns constructed from a symbolic alphabet, generated by the discretization, to quantify the complexity of the dynamics. We adapt this as follows: We compare the positions of $D$ consecutive ensemble elements, assigning to each set of $D$ consecutive elements an ordinal pattern, also known as a {\it{word}}, depending on relative positions of consecutive elements. Specifically, for dimension $D=2$ there are only two words, i.e., $01$ for $x^n<x^{n+1}$, and $10$ for $x^{n+1}<x^{n}$. For dimension $D=3$ there are six possible words, i.e., $012$ for $x^n<x^{n+1}<x^{n+2}$, $021$ for $x^n<x^{n+2}<x^{n+1}$, etc. We then compute a normalized (Shannon) entropy from the probabilities of each word ($p_j$)
\begin{equation}
    \Pi = -\frac{1}{\log(D!)}\sum_{j=1}^{D!} p_j~\log(p_j)
    \label{Pi}
\end{equation}
\noindent where $p_j$ is the probability of the $j$-th ordinal pattern, $D$ the word length or dimension so that $D!$ is the number of possible words of dimension $D$.
This quantifies the loss of spatial correlation at any given time \textbf{relative} to the initial indexed ensemble. In our construction the initial ensemble has the order ($x_1^n < x_1^{n+1},~\forall n$) and therefore yields only one word at $i=1$, whence the entropy is identically zero by construction, independent of the details of the initial $\rho$. However, as we iterate each element, the entropy changes as a function of time. For a chaotic system, for example, we expect an increase up to the limit where all words are equally probable, $\Pi=1$, which defines the range of values for $\Pi$.
%\begin{equation}
 %   PE = -\frac{1}{log(D!)}\sum_{j=1}^{D!} P_j~log(P_j)~,
%\end{equation}
%where $P_j$ is the probability of word $j$. For a purely random process, where all words are equally probable, $PE = 1$, while a process where there is only one word possible, then $PE=0$. PE quantifies the degree of determinism in a time series, and can extract information about the memory of the dynamics.This entropy measure is computationally simple to implement, robust to experimental noise, and a suitable tool to extract temporal correlations in a time series. 
%Note that $\Pi$ is defined considering separations between neighbors of all the members of the ensemble at a given time, and changes in $\Pi$ measure the shuffling inherent in the dynamical folding of $\rho$; this affects local position correlations, but nominally need not depend on the degree of stretching (as measured by $\lambda$).

\begin{figure}[ht]
\centerline{
\includegraphics[width=6.5cm]{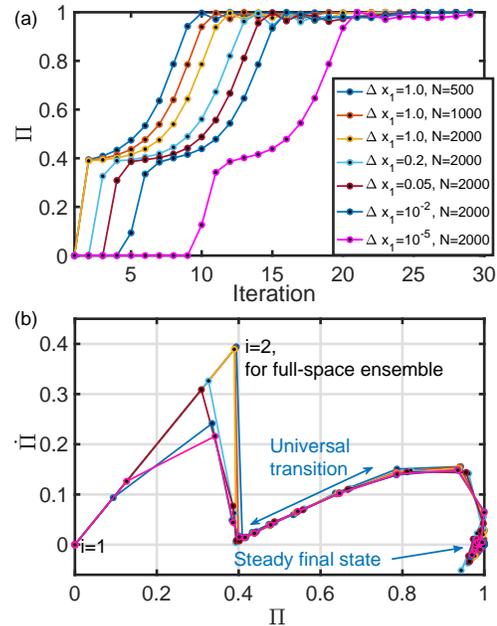}
}
\caption{(a) $\Pi(i)$ for initially equally-spaced ensembles of N elements for the logistic map at $r=4.0$. At $i=1$ the first three ensembles (see legend) cover all of position space ($\Delta x_1=1.0$), but have different $N$ (and density). The last four ensembles have 2000 elements each, but are initially smaller in position space ($\Delta x_1 = 0.2$, 0.05, $10^{-2}$, $10^{-5}$). In all cases $\Pi$ follows an S-shaped curve going from low to high. (b) This behavior mapped in Entropy Phase Space, $[\Pi,\dot{\Pi}]$. The initially fully ordered ensembles start at $[\Pi,\dot{\Pi}]=[0,0]$ and evolve to the right (increasing $\Pi$), and all show a universal transition as the trajectories of the elements braid. Finally the ensembles are fully shuffled with a steady state of maximum entropy, $[\Pi,\dot{\Pi}]=[1,0]$. The initial ensemble is indicated ($i=1$), as well as the position after the first iteration ($i=2$) for ensembles that initially occupy the whole space ($\Delta x_1=1$).}
\label{universal-S}
\end{figure}

In Fig.~\ref{ensemble}(j) we see that after an initial transition period $\Pi(\tilde{\rho})$ follows an S-shaped curve until it reaches the maximum entropy, $\Pi = 1$, precisely as expected for a candidate for thermodynamic entropy~\cite{1999_PRL_Latora, 2002_PLA_Latora, 2003_JETP_Martyushev, 2005_PRE_Falcioni}. This occurs across initial ensembles (Fig.~\ref{universal-S}(a)) where after 
%presents the evolution of Correlation entropy as a function of the iteration, corresponding to ensembles of different number of elements (500, 1000, 2000, 4000). The red plot corresponds to the ensemble from Fig.~\ref{colormap} (N=1000). $\Pi$ is computed with words of dimension $D=3$.
%For the initial conditions, at $n=1$, the distribution of the ensemble is uniformly increasing. This only allows for one ordinal pattern in the whole distribution, $012$ ($x_{i-1} < x_i < x_{i+1}$), therefore $A=0$. The initial uniformity is broken at iteration $n=2$. This induces a sharp increase in $A$ which comes from having now a distribution with two possible words of equivalent probability (012 and 210). This leads to $A \approx -\frac{2\times \frac{1}{2}log(\frac{1}{2})}{log(6)} \approx 0.387$. This can be appreciated in Fig.~\ref{colormap}, where $n=2$ shows how the distribution grows for $1 < i < 500$ and decreases for $500 < i <1000$ (blue on the sides and yellow in the middle).
an initial-$\rho$-dependent `pre-thermalization' transient stage there is universal behavior. The time scale for the onset of the second stage increases when the sampling density increases, as is intuitive. The universality of these dynamics including the characteristic relaxation timescales is readily visible in the parametric  $(\Pi_i,\dot{\Pi_i})$ plot in Fig.~\ref{universal-S}b for all the ensembles from Fig.~\ref{universal-S}a. In this {\it{Entropy Phase Space}} (EPS) each ensemble evolves along a different pre-thermalization trajectory, and they all converge for the final linear transition to a high-entropy steady state.
% $n=1$ corresponds to the fully ordered initial state of the ensembles ($\Pi=0$, $\dot{A\Pi}=0$). Then the entropy increases and shows a high slope ($\Pi \approx 0.4$, $\dot{\Pi}\approx 0.4$). The universal transition takes place after that iteration (n=2) and lasts until the ensemble reaches maximum entropy ($\Pi \approx 1$, $\dot{\Pi} \approx 0$).
We fit this linear trajectory for $(\dot{\Pi_i}, \Pi_i)$ as $ \Pi_i=\Pi_0e^{\alpha i}$, where $i$ indicates the iteration. The exponent $\alpha$ measures the growth rate of $\Pi$, using folding in phase space as a measure. Despite not being explicitly constructed using stretching rates, the loss of information due to folding should arguably relate to the information loss rate for the dynamics, i.e. to $\lambda$ or the PE itself. 

\begin{figure}[ht]
\centerline{
\includegraphics[width=9.0cm]{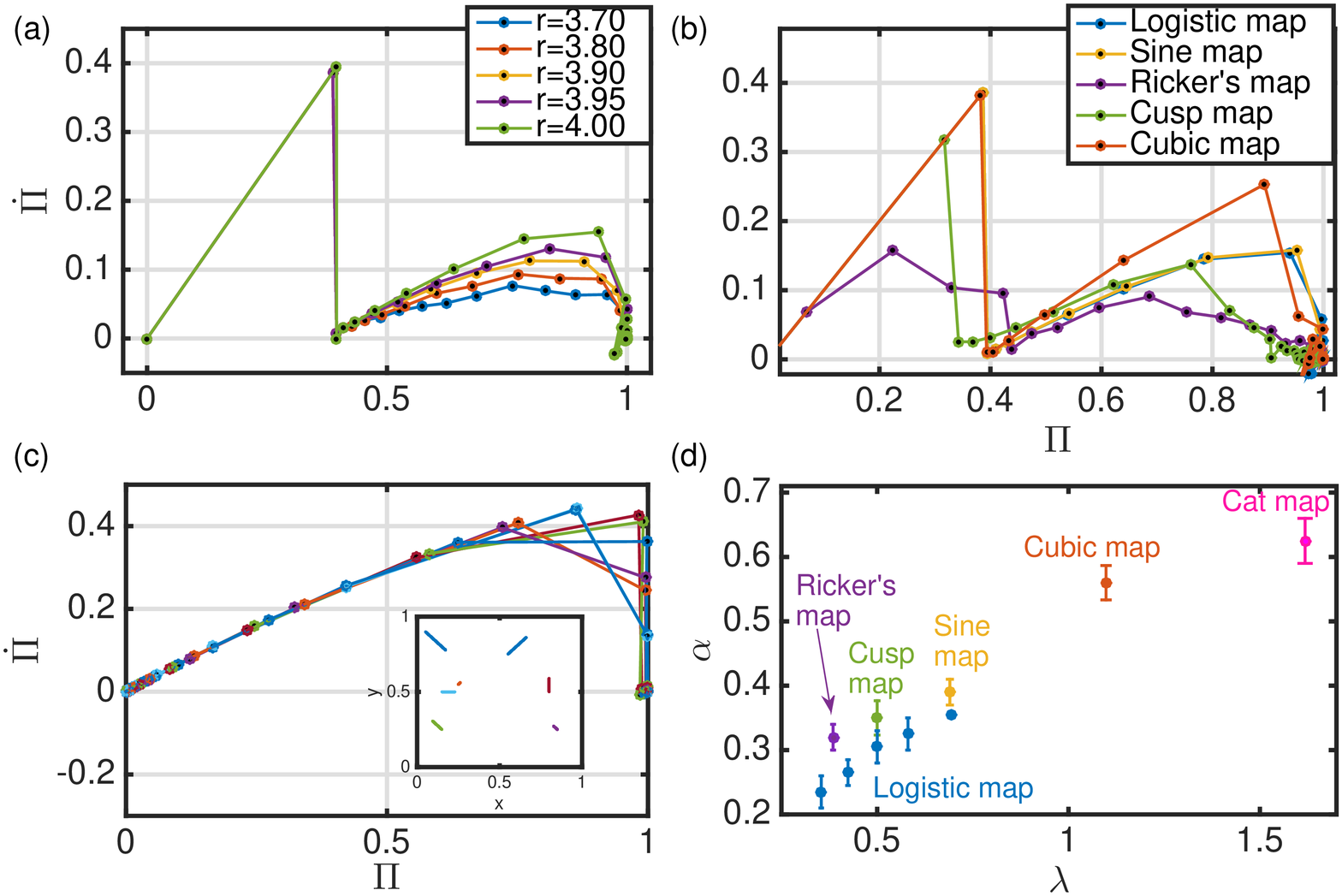}
}
\caption{Entropy Phase Space (EPS) for (a) the logistic map with varying degrees of chaos ($r=3.7, 3.8, 3.9, 3.95, 4.0$).
The initial ensemble $\rho$ (and $\tilde{\rho}$) is the same for each case (N=1000, $\Delta x^i_1 = 10^{-3}$). All show the universal S-shaped transition where the higher $r$ (and Lyapunov exponent $\lambda$) correspond to a larger exponent $\alpha$.
(b) EPS for the Logistic map (r=4, $\lambda=0.69$), Cubic map ($x_{i+1,n}=rx_{i,n}(1-x_{i,n}^2)$, r=3.0, $\lambda=1.1$), Sine map ($x^{i+1,n}=r\sin(\pi x_{i,n})$, r=1, $\lambda=0.69$),
Ricker's map ($x_{i+1,n}=rx_{i,n}e^{-x_{i,n}}$, r=40, $\lambda=0.39$), and Cusp map ($x_{i+1,n}=1-r\sqrt{|x_{i,n}|}$, r=2, $\lambda=0.5$).
(c) EPS for the Cat map for different initial $\rho$, showing the same universality found in 1-D. Each curve corresponds to one of the initial spatial distributions of $\rho$ shown in the inset.
(d) $\alpha$ versus $\lambda$.
}
\label{EPS}
\end{figure}

When the degree of chaos is changed using $r$, as shown in Fig.~\ref{EPS}(a) we find that $\alpha$ changes monotonically as $r$ changes from $r=3.7$ to $r=4.0$ in the fully chaotic regime for this system.
%This behavior is quantified through the Correlation Exponent, extracted from the slope of the universal transition in EPS (Fig.~\ref{alpha}b). 
All of these behaviors prove to be generic, and not unique to the logistic map. Figure~\ref{EPS}b shows that the same results (S-shaped thermalization transition, a universal linear stage in the EPS, and an increase of $\alpha$ with $\lambda$) are obtained in various other chaotic 1-D iterative maps for control parameters with a range of dynamics and Lyapunov exponents. 

%\section{Arnold's cat map.}

These useful properties of $\Pi$ generalize and scale well in computational difficulty to 2-D systems where we consider in particular those drawn from time-dependent Hamiltonian dynamics. We start with the area-preserving uniformly hyperbolic two-dimensional stretching and folding dynamics of the Arnold's Cat Map 
\begin{equation}
\begin{cases}
        x_{i+1} = x_i+y_i~;~\mod(1)\\
        y_{i+1} = x_i+2y_i~;~\mod(1).
\end{cases}
\label{eq_cat}
\end{equation}
%We consider an ensemble of N elements distributed in $[x,y]$-space and study its time evolution by iterating Eqs.~\ref{eq_cat}. In order to explore how a highly structured, low entropy, ensemble evolves to a high entropy state, we choose an initially ordered ensemble.
Figure~\ref{EPS}(c) shows the EPS for the Cat map using separations in $x$ (equivalent results are found using $y$). The various $\rho$ are initialized as different lines in phase space (see inset) with $\Pi=0$. All show the same S-shaped evolution (see Supplementary Information) as for 1-D chaotic maps, and the EPS in Fig.~\ref{EPS}(c) shows trajectories converging to a final linear stage with $\alpha=0.63$. In Fig.~\ref{EPS}(d) we see that $\alpha$ in fact is monotonically though nonlinearly correlated with $\lambda$ across all these dynamical systems.

%\begin{figure}[htbp]
%\centerline{
%\includegraphics[width=6.5cm]{figures/cat_map.eps}
%}
%\caption{(a) Evolution of $\Pi$ for the Cat map for different initial ensembles. Entropy rate shows the same S-shaped curve found in the one-dimensional maps. Each curve corresponds to one of the initial $\rho$ shown in the inset. This represents the initial spatial distribution of $\rho$. (b) Entropy phase space. All $\rho$ follow the same universal path and share the same slope.}
%\label{cat_map}
%\end{figure}
%\section{The Standard map.}

The Chirikov Standard Map is the 2-D area-preserving Map 
\begin{equation}
\begin{cases}
    p_{i+1} = p_i+K~\sin(\theta_i)~;~mod(2\pi)\\
    \theta_{i+1} = \theta_i+p_{i+1}~;~mod(2\pi)
    \end{cases}
\end{equation}
with dynamics constrained to $0\leq [\theta,p] \leq 2\pi$, and where $K$ is the nonlinear kick strength. For $K=0$ the system is linear and the dynamics periodic; as $K>0$ increases, the dynamics can be chaotic or regular depending on initial conditions, unlike the uniformly hyperbolic Cat Map. In general both the chaotic fraction of phase space and $\lambda$ increase with $K$. The Standard Map's `mixed' phase space, which is expected for generic Hamiltonians, leads to a challenging complexity of behavior~\cite{2017_JSMTE_Tsallis} fundamental to understanding non-equilibrium thermodynamic phenomena such as non-equilibrium steady-states. 

\begin{figure}[ht]
\centerline{
\includegraphics[width=9.5cm]{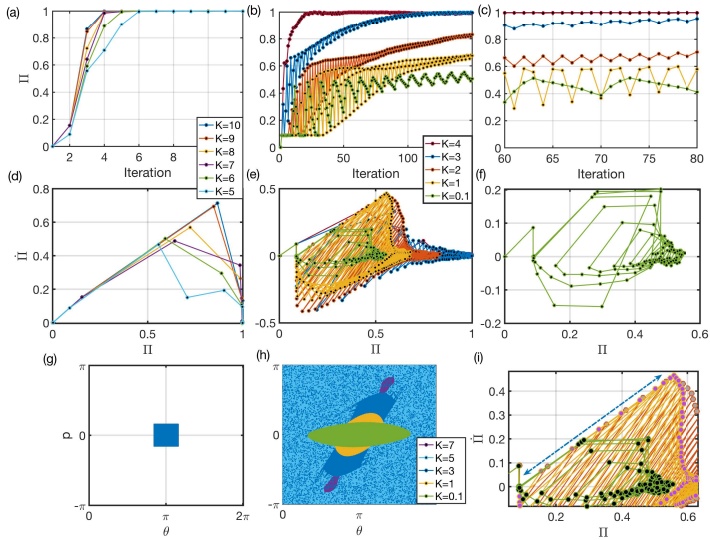}
}
\caption{(a) Evolution of $\Pi$ for the Standard map for $5 \leq K \leq 10$. Each ensemble $\rho$ is computed with 4096 elements, occupying the central region (g). $\Pi$ grows faster as $K$ and therefore $\lambda$ increases. (d) EPS for (a). (b) Evolution of $\Pi$ for the Standard map for $0.1 \leq K \leq 4$. As $\Pi$ grows it shows oscillations, signature of the mixed dynamics in $\rho$. (e) EPS for (b). The envelope unveils the expected linear relation. (g) Initial $\rho$ in phase space. (h) Phase space occupied by $\rho$ as it evolves (for different values of $K$). For $K=7$ (purple) $\rho$ finally occupies the whole phase space, but the figure only shows the region that is not overlapped by the other ensembles. For $K=5$ $\rho$ evolves to occupy most of the phase space, except for a narrow central region, that allows to see the purple $\rho$ for $K=7$. The smaller the chaotic contribution (lower $K$) the more localized the evolution of the $\rho$. (c) Detail of $\Pi$ from (b) where one can observe the structure of $\Pi$'s evolution and some of the frequencies involved for each $K$. (f) EPS for $K=0.1$ where the oscillations are presented as a spiraling down to equilibrium ($\Pi=1$). (i) Detail of the EPS for $K=0.1,~ 1.0,~ 2.0$ that highlights how they share the linear envelope.}
\label{standard_1}
\end{figure}

Recent work~\cite{2017_JSMTE_Tsallis} has shown that for large enough $K$ the system relaxes to equilibrium with dynamics like a uniformly hyperbolic system. For small $K$ they see similar relaxation to equilibrium along with (limited) evidence of entropy oscillations for initially sharply localized states. Their computations are not computationally atypical, using $10^5$-$10^6$ trajectories and some novel measures (SALI) for characterizing thermalization which unfortunately do not generalize as a function of $K$. We find that using $\Pi$ entropy allows us to push beyond these limits. Figure~\ref{standard_1} shows $\Pi$-Entropy dynamics for sharply localized initial ensembles (Fig.~\ref{standard_1}(g)) for the range $0.1\leq K \leq 10$. We see that the evolution and `final' (on the scales of our study) state depend on $K$ (Fig.~\ref{standard_1}(h)) in clearly distinguishable and informative ways. 

Specifically, for large $K$, where chaos dominates and the system is essentially uniformly hyperbolic, $\Pi$ indeed follows an S-shaped curve (Fig.~\ref{standard_1}(a,d)) in agreement with the previous results~\cite{2017_JSMTE_Tsallis}. At lower $K$ (Fig.~\ref{standard_1}(b,e)) we find that $\Pi$ reveals the richness of thermalization dynamics in the mixed phase of the dynamics, visible in the rapid oscillations overlaid with complex envelopes. These structures arise from the way that initial ensembles include trajectories from different dynamical regimes, but we see in all cases a final similar saturation (Fig.~\ref{standard_1}c). In EPS these mixed phase space trajectories evolve as spirals that drift increasingly more prominently with $K$ to the right and end with a final linear steady state (Fig.~\ref{standard_1}(f,i)).

\begin{figure}[htbp]
\centerline{
\includegraphics[width=9cm]{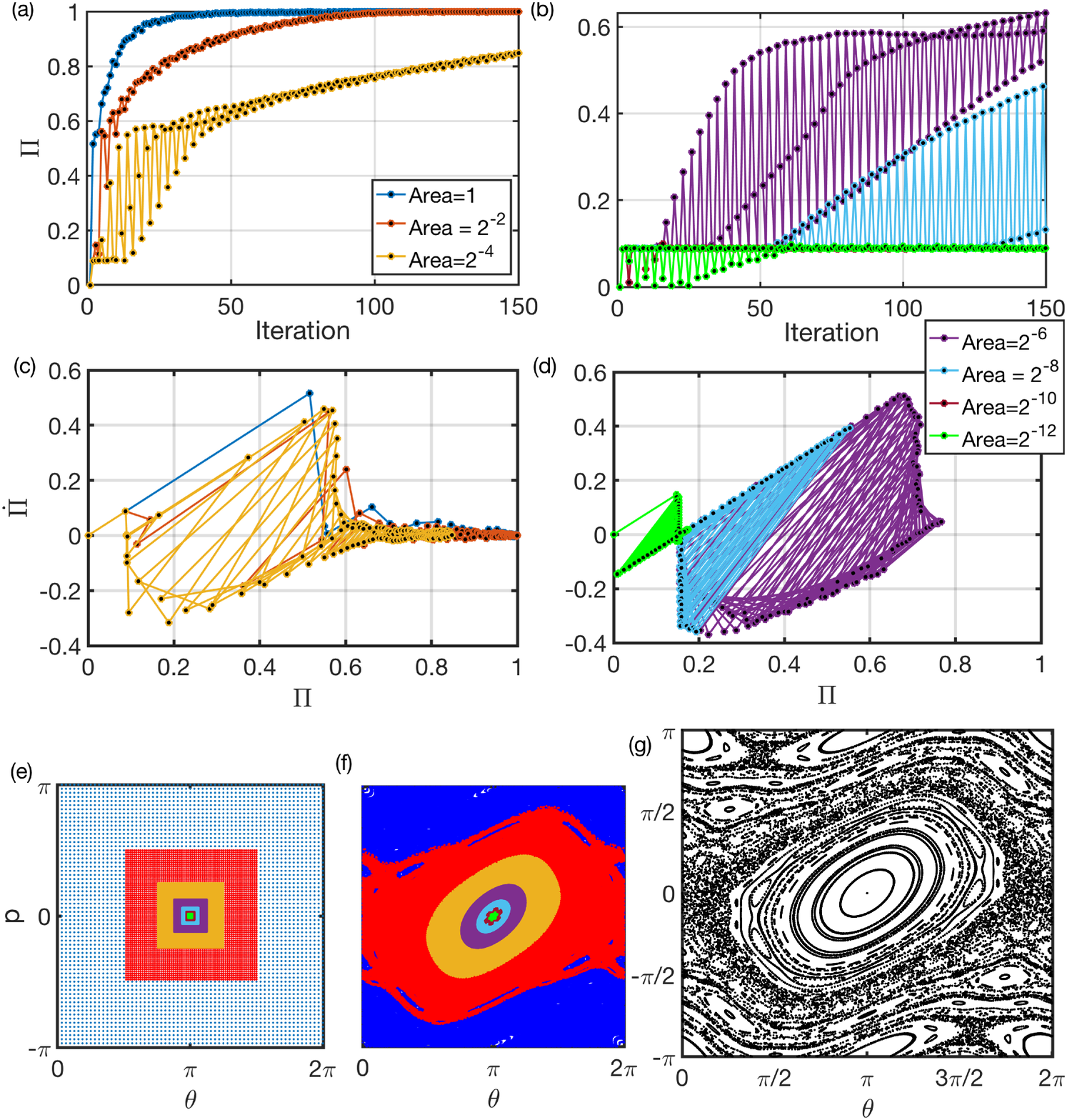}
}
\caption{(a) Evolution of $\Pi$-entropy for the Standard map for $K=1.0$. Each ensemble $\rho$ is computed with 4096 elements and corresponds to a different initial area in phase space. (b) Same as (a) but for $\rho$ occupying smaller areas in phase space.
(c) Corresponding EPS for (a). (d) Corresponding EPS for (b). $\rho$ corresponding to an area $A=2^{-10}$ overlaps the $\rho$ of initial area $10^{-12}$. Although the behavior in phase space is oscillatory, the envelope of the evolution of each $\rho$ shows the same slope corresponding to the relaxation. (e) Initial region occupied by each $\rho$, all centered at $(\theta = \pi,p=0)$. (f) Region explored by each $\rho$ in their evolution. (g) Phase space of the standard map for K=1.0.}
\label{standard_2}
\end{figure}

We unpack how initial $\rho$ localization affects the thermalization in Fig.~\ref{standard_2} where we use $K=1.0$, a parameter where there exist a substantial fraction of both periodic and chaotic regions (Fig.~\ref{standard_2}(g)). All results shown consider $\rho$ centered at $(\pi,0)$ (a fixed point of the dynamics), with sizes ranging from the entire phase space (area$~1$, blue ensemble), to a microscopic initial state (area $= 10^{-12}$, green ensemble), and show results from the first $150$ iterations. The most localized $\rho$ contain periodic trajectories only and this periodicity is visible in both the $\Pi$ dynamics and the EPS. As the number of chaotic trajectories within a sample increases, $\Pi$ dynamics gets increasingly complex, showing multi-frequency oscillatory growth.

\begin{figure}[htbp]
\centerline{
\includegraphics[width=9cm]{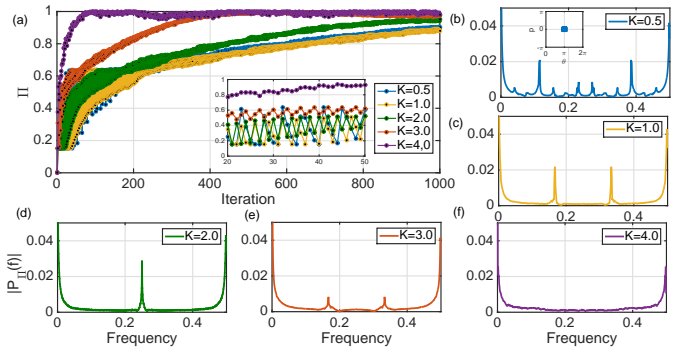}
}
\caption{(a) Evolution of $\Pi$-entropy for the Standard map for 1000 iterations for several $K$ values. Inset highlights the oscillatory behaviour shown within the overall growth of $\Pi$. (b-f) Fast Fourier Transform of the temporal evolution of $\Pi$ for $K=0.5, 1.0, 2.0, 3.0, 4.0$. Inset on (b) indicates the initial ensemble. Different frequencies are manifest for different $K$ values. For $K=4$ there are no dominant frequencies.}
\label{standard_FFT}
\end{figure}

In EPS, we see that the different entropy trajectories do not overlap (the changing weight of the periodic fraction changes the thermalization strongly) but have similar envelopes. 
Fourier analysis of $\Pi$ dynamics (Fig.~\ref{standard_FFT}) shows how $K$ strongly affects dominant frequencies. For $K=0.5$, there is a rich frequency spectrum with clear peaks (Fig.~\ref{standard_FFT}(b)), while $K=4.0$ shows a flatter and more typically chaotic spectrum (Fig.~\ref{standard_FFT}(f)). To capture the linear regime in the EPS along with this complex behavior across all these dynamics, we model the dynamics using $\Pi = (1 -\exp(-\alpha t))[A+ B \cos(\omega t)\exp(-\alpha t)]$, where $\alpha$ is the relaxation exponent, $\omega$ a characteristic oscillation frequency, and $A$ and $B$ estimate the relative support of the initial distribution in the chaotic regime and regular regime respectively.

All of these results suggest that $\Pi(\tilde{\rho})$ is indeed able to distinguish a variety of macroscropic time scales that depend on the system's microscopic dynamics, in particular for the global relaxation of an ensemble to equilibrium across a variety of situations. We believe that the PI-entropy can be implemented advantageously in experimental situations since it requires tracking a comparatively small number $(100 -1000)$ of individual trajectories in an ensemble, well within the reach of experiments using tracer particles\cite{2021_PNAS, 2019_NatComm, 2020_PRL_cells}. Further, while the details of this technique are here presented in the context of non-interacting ensembles for dynamical maps, the method is general and can be extended to a variety of complex dynamical systems where ensembles are transitioning from a low entropy to a high entropy state. We remark that this method captures both small-scale and large-scale correlations and structure and has built-in coarse graining. It seems thus to discard non-essential aspects of the ensemble dynamics as an alternative to constructs such as the Boltzmann-Gibbs Entropy. We envision this to prove a powerful tool in understanding the evolution and loss of correlations in complex systems as varied as the mobility of individuals, fluid dynamics, solitons, many-body localization, etc; in short, for any multi-element dynamical system that evolves from an ordered distribution state to a disordered one.

%\begin{acknowledgments}
%We wish to acknowledge the support of the author community in using
%REV\TeX{}, offering suggestions and encouragement, testing new versions,
%\end{acknowledgments}

\nocite{*}

\bibliography{apssamp}% Produces the bibliography via BibTeX.

\end{document}